# Sigma-Lognormal Modeling of Speech

C. Carmona-Duarte[1] · M. A. Ferrer[1] · R. Plamondon[2] · A. Gómez-Rodellar[3] · P. Gómez-Vilda[3]

**Abstract**
Human movement studies and analyses have been fundamental in many scientific domains, ranging from neuroscience to education, pattern recognition to robotics, health care to sports, and beyond. Previous speech motor models were proposed to understand how speech movement is produced and how the resulting speech varies when some parameters are changed. However, the inverse approach, in which the muscular response parameters and the subject's age are derived from real continuous speech, is not possible with such models. Instead, in the handwriting field, the kinematic theory of rapid human movements and its associated Sigma-lognormal model have been applied successfully to obtain the muscular response parameters. This work presents a speech kinematics-based model that can be used to study, analyze, and reconstruct complex speech kinematics in a simplified manner. A method based on the kinematic theory of rapid human movements and its associated Sigma-lognormal model are applied to describe and to parameterize the asymptotic impulse response of the neuromuscular networks involved in speech as a response to a neuromotor command. The method used to carry out transformations from formants to a movement observation is also presented. Experiments carried out with the (English) VTR-TIMIT database and the (German) Saarbrucken Voice Database, including people of different ages, with and without laryngeal pathologies, corroborate the link between the extracted parameters and aging, on the one hand, and the proportion between the first and second formants required in applying the kinematic theory of rapid human movements, on the other. The results should drive innovative developments in the modeling and understanding of speech kinematics.

✉ C. Carmona-Duarte
ccarmona@idetic.eu

M. A. Ferrer
mferrer@idetic.eu

R. Plamondon
rejean.plamondon@polymtl.ca

A. Gómez-Rodellar
andres.gomez@ctb.upm.es

P. Gómez-Vilda
pedro@fi.upm.es

[1] Instituto Universitario Para El Desarrollo Tecnológico Y La Innovación en Comunicaciones, Universidad de Las Palmas de Gran Canaria, Las Palmas de Gran Canaria, Spain

[2] Laboratoire Scribens, Département de Génie Électrique, Polytechnique Montréal, Montreal, QC, Canada

[3] Facultad de Informática, Universidad Politécnica de Madrid, Campus de Monte-Gancedo, s/n, 28660 Boadilla del Monte, Madrid, Spain

## Introduction

For decades, human movement studies and analyses have been fundamental in many scientific domains, ranging from neuroscience to education, pattern recognition to robotics, health care to sports, and beyond. The primary goal of these studies has always been to parameterize and assess human movements, providing information on the basic processes involved in fine motor control and their variability. In speech, computational systems that synthesize and assess speech motor control provide answers to some questions regarding the articulator movements used by humans to produce speech sounds, speech rate effects, or for example, how infants acquire the motor skills needed to produce the speech sounds of their native language [1]. However, many questions regarding the modeling and automatic assessment of natural neuromotor decline in healthy speech or the parameterization of neuromotor commands and muscular responses from fast continuous speech are still open.

Many neurocomputational models have been proposed to understand how speech movement is produced and how the resulting speech varies when changing some parameters [2]. To this end, motor control models inspired by computer programs have often been used. Under this paradigm, motor commands are generated based on a central motor plan [2] and executed by a speech generator. Several speech motor models, such as the GEPETTO [3, 4], ACT [5], DIVA [6] and Task Dynamics [7], State Feedback [8], and FACTS [9] models, have been developed in this context in recent years. These models start from an action plan (planner) and then adjust a set of parameters, moving a set of articulators, to get the ideal output (feedforward models). In some of them, the acoustic output signal is then compared with the reference input signal from the planner to generate an error signal that allows to correct the movement (feedback models).

Previous works have been oriented toward the modeling of learned speech of healthy speakers. Some of the above models (DIVA, FACTS, and ACT) can, following adjustments of some parameters, model certain aspects of development and aging. However, to the best of our knowledge, neuromotor decline has never been modeled by such systems [2]. Moreover, the inverse approach, in which the muscular response parameters are derived from real continuous speech, is not possible with such models.

Among the models that study human movement production in general [10], the kinematic theory of rapid human movements and its associated Sigma-lognormal model [11–13] have been applied successfully in several fields [14] to model numerous human movements such as handwriting and signatures, as well as eye, finger, wrist, hand, head, and trunk movements [14–17]. It has also been used to evaluate the effect of exercise on global neuromotor control [18], on the detection and monitoring of neuromuscular disorders, and to study and synthesize handwriting motor control changes in humans with age [19–21]. The Sigma-lognormal model has thus demonstrated its capacity to obtain a muscular response and neuromotor command parameters from online handwriting, to assess neuromotor aging and synthesize new handwriting samples.

In handwriting, the Sigma-lognormal model decomposes a complex movement, obtained from the temporal trajectory captured with a digital tablet, into a sum of simple time-overlapped primitives with a lognormal velocity profile. This method provides information about how every single movement is generated and synchronized, modeling the end effector (set of muscles involved in the movement) as a black box. Thus, the lognormal-shaped impulse response of the end effector, used as a primitive, is not linked to any specific articulation, but rather, to a large number of coupled subsystems. Moreover, the movement primitive is not necessarily confined to movements with a single velocity peak, as is still often assumed in many models [22].

Given the above advantages, in this paper, we propose a novel methodology based on the Sigma-lognormal model to parameterize the speech kinematics and the muscular response produced by the complex set of muscles involved in achieving the target sound, as well as to study aging effects. One question that does arise though is how the kinematic theory can be applied to speech modeling. The answer to this question is by no means straightforward. As a first proof of concept, preliminary works directly applied the kinematic theory of rapid human movements to diphthongs and sustained vowels uttered in neuromotor disease analysis [23–26], suggesting the possibility of applying the Sigma-lognormal model to speech. However, obtaining a general model would require a representation of a target's map, a trajectory mapping, and a velocity representation, all assuming a lognormal impulse response that would need to be related to some speech features. To address these issues, we assume a high-level goal as the target map (a map of sound that can be discriminated between them), inspired by the work on the spatial model proposed by Moser et al. [27, 28], instead of a fixed desired position of each individual speech articulator. As such, the velocity representation can be obtained from the sound transitions (trajectory map). The model is explained in detail in "Sigma-Lognormal Parameterization Method".

To test the validity of the proposed method, we present two sets of experiments. The first one aims to illustrate the meaning of the lognormal decomposition in simple movements in a continuous speech signal. In the second one, the goal is to evaluate the model's ability to identify significant differences in some parameters when modeling aging in the speech of subjects with or without laryngeal pathologies. In certain studies related to handwriting [19], it has been observed that the time between lognormals and their number increases with age. Timing effects have also been reported in speech, where an fMRI study suggested that the motor control of timing during speech production declines with age [29]. So, if the proposed Sigma-lognormal model describes the speech kinematics well, then we should expect results obtained in speech to be similar to those obtained in handwriting [19] if proper experiments are run. Moreover, since laryngeal dysfunction only affects the sound source (glottis), and not the global end effector movements, the time between lognormals should not be affected in this case, unlike in the case of aging. In the experiment section, these hypotheses will also be tested.

The present work is structured as follows. After an overview of the kinematic theory of rapid human movements in "Overview of the Sigma-Lognormal Model", "Sigma-Lognormal Parameterization Method" describes the method for estimating speech kinematics and how it is parameterized. "Evaluation, Results, and Discussion" evaluates the model and discusses the results obtained. Finally, we summarize our findings in "Conclusions".

## Overview of the Sigma-Lognormal Model

The Sigma-lognormal model explains how an action plan comprised a sequence of circumference arcs between virtual target points (VTP) can be activated to generate a spatiotemporal trajectory. Virtual target points are defined as the positions targeted by a lognormal, but that are not necessarily reached because of the temporal overlapping of the next lognormal [30]. Virtual targets are thus related to the learning process and how the movement is programmed by the brain. A starting and an ending angle define each arc linking virtual target points. Each ending VTP is the starting VTP of the next arc. To generate smooth movements from this discontinuous action plan, the instantiation of a command at a given VTP must start before the previous stroke reaches that VTP. In other words, each arc has a starting time but finishes later than the starting time of the next one. Therefore, successive resulting strokes are temporally overlapped. Each arc is executed following a lognormal-shaped velocity curve, and the whole trajectory is made up of the vector summation of the individual strokes.

Mathematically, the lognormal velocity profile of a simple movement is defined by [7]

$$\left|\vec{v}_j(t;t_{oj})\right| = D_j \Lambda_j(t;t_{oj},\mu_j,\sigma_j) = \frac{D_j}{\sigma_j \sqrt{2\pi}(t-t_{oj})} e^{-\frac{(\ln(t-t_{oj})-\mu_j)^2}{2\sigma_j^2}} \quad (1)$$

where $D_j$ is the length of the movement, $t_{oj}$ is the time occurrence of the movement command, $\mu_j$ is the log time delay, $\sigma_j$ is the log response time, and $j$ indicates the index of the movement. The velocity profile of a complex movement $\vec{v}_r(t)$ is given by the time superposition of $NbLog$ lognormals [9] as follows:

$$\vec{v}_r(t) = \sum_{j=1}^{NbLog} \vec{v}_j(t) = \sum_{j=1}^{NbLog} D_j(t) \begin{bmatrix} \cos\phi_j(t) \\ \sin\phi_j(t) \end{bmatrix} \Lambda_j(t;t_{oj},\mu_j,\sigma_j) \quad (2)$$

where $\phi_j(t)$ is the angular position, defined as

$$\phi_j(t) = \Theta_{sj}(t) + \frac{\Theta_{ej}(t) - \Theta_{sj}(t)}{2} \left[ 1 + erf\left(\frac{\ln(t-t_{oj}) - \mu_j}{\sigma_j\sqrt{2}}\right)\right] \quad (3)$$

and $\Theta_{sj}(t)$ and $\Theta_{ej}(t)$ are the starting and the end angular directions of the $j^{\text{th}}$ simple movement or stroke, and $erf$ is the error function.

Finally, the trajectory is worked out as

$$\vec{s}_r(t) = \begin{bmatrix} x_r(t) \\ y_r(t) \end{bmatrix} = \sum_{j=1}^{NbLog} \frac{D_j}{\Theta_{ej} - \Theta_{sj}} \begin{bmatrix} \sin\phi_j(t) - \sin\Theta_{sj} \\ -\cos\phi_j(t) + \cos\Theta_{sj} \end{bmatrix} \quad (4)$$

This expression converts angles into arcs of circumferences that are temporally overlapped. Specifically, the $j^{\text{th}}$ term of the summation represents the arc that links consecutive virtual target points, $VTP_{j-1}$ and $VTP_j$, which are defined by

$$VTP_j = VTP_{j-1} + \frac{D_j}{\Theta_{ej} - \Theta_{sj}} \begin{bmatrix} \sin\phi_j(T) - \sin\Theta_{sj} \\ -\cos\phi_j(T) + \cos\Theta_{sj} \end{bmatrix} \quad (5)$$

with $T$ being the total temporal duration of the spatiotemporal sequence.

A sequence of virtual target points, along with their starting and ending angles and their lognormal velocity parameters, can be analytically extracted through reverse engineering (Fig. 1). Using the extracted action plan, the corresponding spatiotemporal sequence can be reconstructed from its set of parameters:

$$P = \{D_j, t_{oj}, \mu_j, \sigma_j, \Theta_{ej}, \Theta_{sj}, VTP_{j-1}\}_{j=1}^{NbLog} \quad (6)$$

Classically, these parameters are calculated from the sampled 2D spatiotemporal sequence with software such as ScriptStudio [31] or iDeLog [32].

Once the original velocity $v_o(t)$ has been reconstructed as a summation of lognormals ($\vec{v}_r(t)$), the quality of the reconstruction can be evaluated using the signal-to-noise-ratio (SNR) between them. Specifically, the SNR is defined as [30]

$$SNR = 20\log\left(\frac{\int_0^T v_o^2(t)}{\int_0^T |v_o(t) - v_r(t)|^2 dt}\right) \quad (7)$$

It is commonly accepted that when SNR < 15 dB, the reconstruction is not appropriate due to either ScriptStudio [31] or iDeLog [32] not having managed to find an adequate solution or to the spatiotemporal sequence not corresponding to the model [30]. In the latter case, as the lognormal is accepted as a neuromotor model, we could also say that the spatiotemporal sequence does not correspond to the timing conditions under which lognormals emerge, as predicted by the central limit theorem [33].

## Sigma-Lognormal Parameterization Method

The scheme for applying the Sigma-lognormal model in speech is presented in Fig. 2. The model divides the speech generation into two steps: planning of the sequence of sounds (effector-independent) and execution

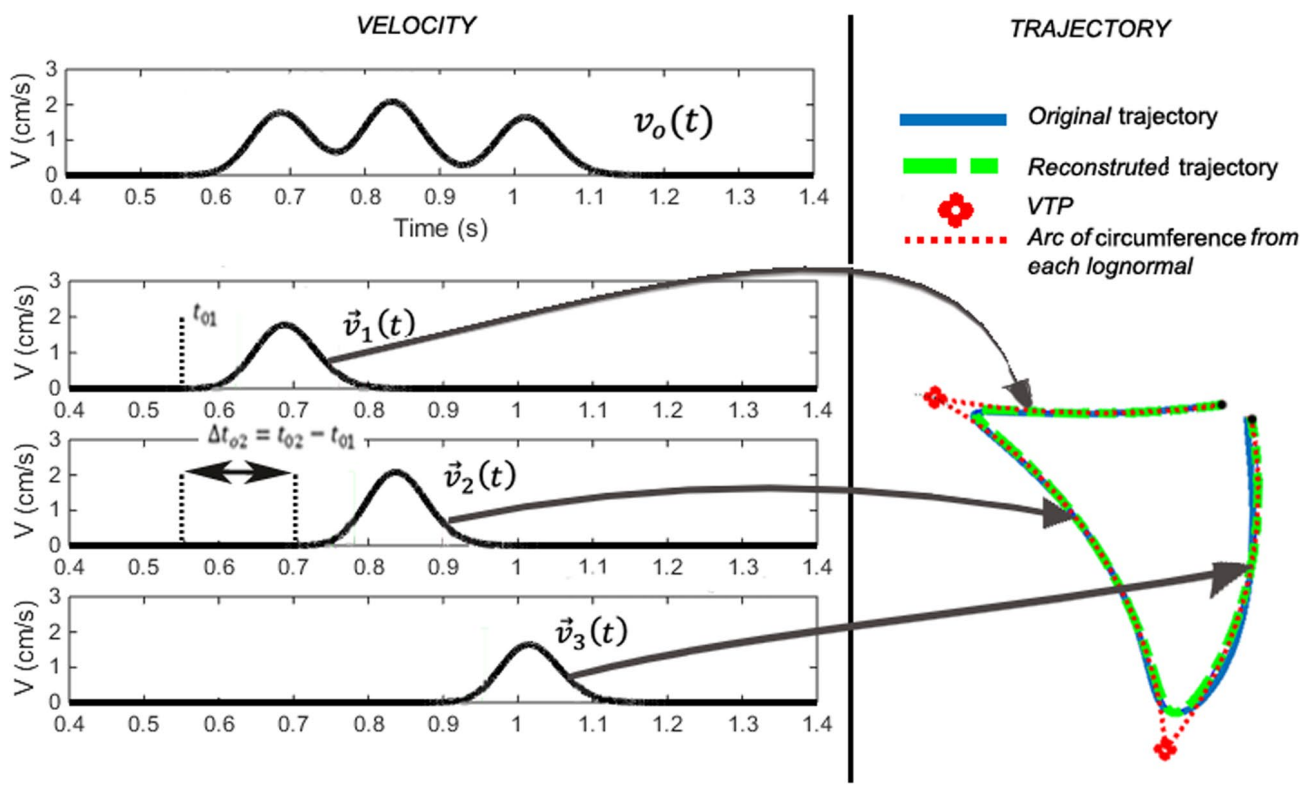

**Fig. 1** Sigma-lognormal reverse engineering of a signal. Decomposition of the velocity profile into a sum of lognormals and the analyzed trajectory for each lognormal

of the sequence via the end effector (effector-dependent) [20, 21, 34]. Firstly, in the effector-independent step, a sound map (higher-level goal) is defined, assuming that each simple learned sound has a corresponding position on a hexagonal grid. Note that in this map (different for each person), the targets are sounds, and not phonemes, since a phoneme can be defined either as a simple sound or as a group of different sounds. Processing a sequence of sounds (for example, [uiau] in (Fig. 2) involves moving through different positions on the grid and generating a trajectory through the selected sounds from a series of commands. Secondly, the effector-dependent module is linked to the neuromuscular system itself (end effector) and is defined by its impulse response to each command. The end effector movement causes the vocal track shape to vary, thus changing the resonance frequencies, and therefore, the formants (resonant frequencies of the oronasopharyngeal tract) over time [35–37].

In this work, we use a reverse engineering approach, starting from the formants and moving up to the sound trajectory; from variations in the formant, we estimate both the parameters that model the commands that determine the transition from one grid position to another (simple movement) and the muscular response to each command. Thus, we will model neither the initial positions on the grid nor the physical constraints of the vocal track.

To perform a Sigma-lognormal analysis of speech, a spatiotemporal sequence that globally represents the speech kinematics is required, as previously depicted in Fig. 2. To this end, we rely on the resonance tube model. Indeed, in speech synthesis, the vocal tract (from the glottis to lips) can be represented as a concatenation of lossless acoustic tubes, where the shape and the volume of the vocal tract vary for each sound [36, 38]. An increment or decrement of the section and length of the tubes produces a change in the resonance frequencies, and accordingly, a change of formants in the output speech, as we can see in Fig. 3. This means that each motor command that the brain produces to generate synchronous muscle movement required to go from one acoustic position to another changes the resonant cavities. Thus, a relationship can be established between an increment of the formant and the increment of the resonant areas or between the formant tracks and the movements of muscles [39]. Then, if the estimated velocity is integrated, a kinematic trajectory to be analyzed by our model can be obtained. Note that the resonant cavities of each subject are different, depending on the morphology and length of the organs that comprise it. Therefore, when a language is learned, the articulatory position of each sound is set as a function of the resonant cavities needed to produce the sound closest to the ideal one the person is trying to learn, and of how the sound is perceived [40–42].

**Fig. 2** Scheme of the proposed model

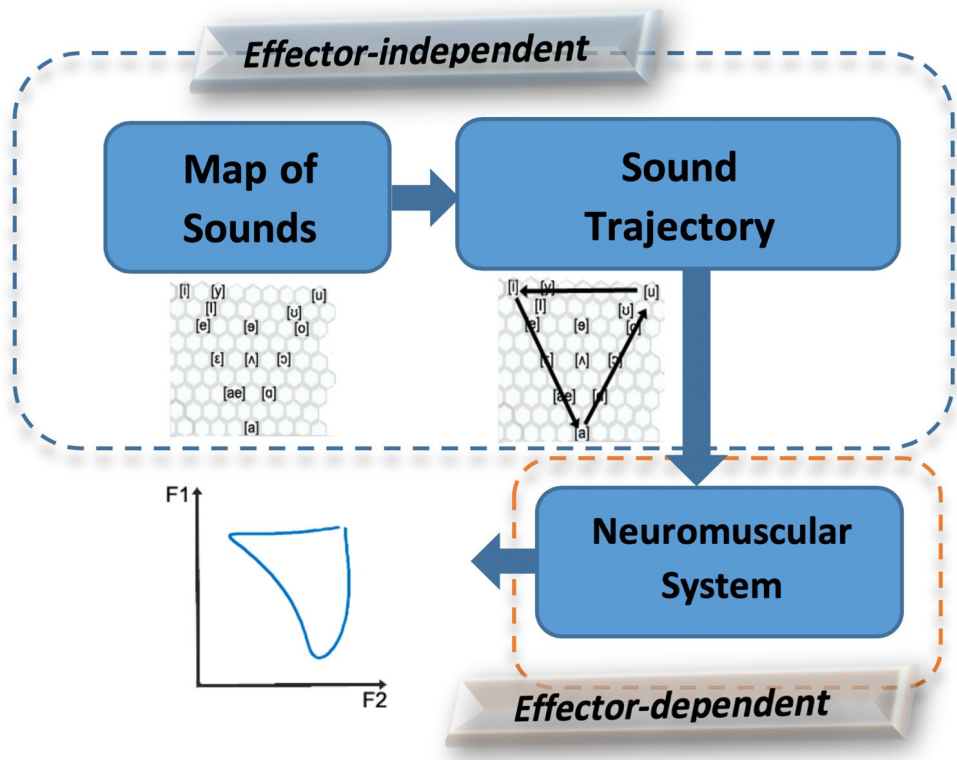

This kinematic trajectory of the formants can be considered as the movement of a reference center (RC) of a speech end effector over the acoustic space, much the same as the movement of the pencil tip over paper represents the movement produced by the end effector during handwriting.

**Fig. 3** Variations in the areas of the tube with time (left) produce a change in formant space (right)

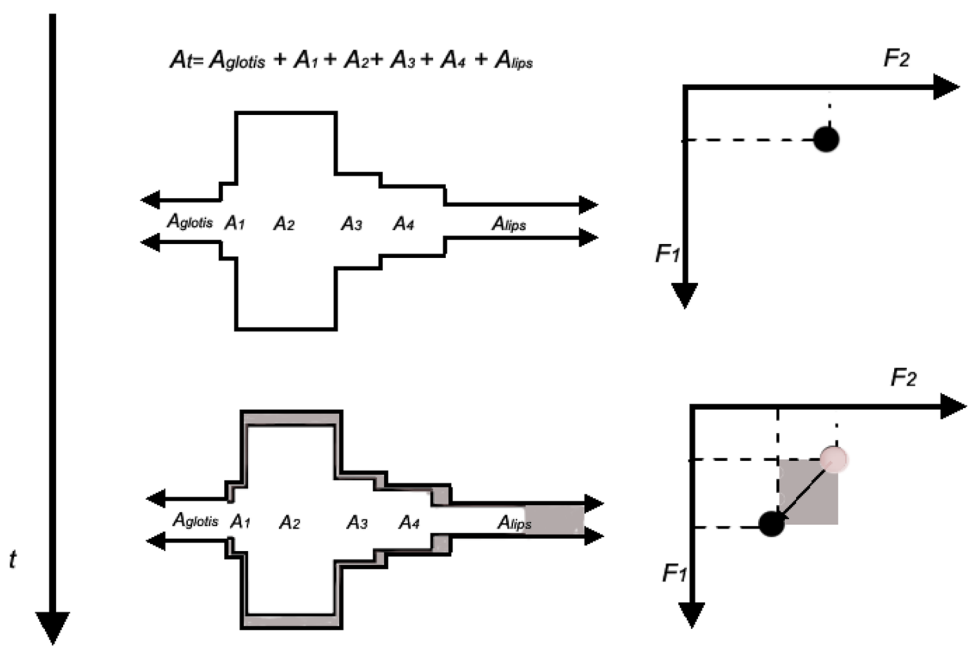

According to this analogy, the model parameters could be recovered from a speech signal in three steps: (1) tracking of the formants; (2) from the formant sequence, obtention of the end effector kinematics, and (3) parameterization of the resulting trajectory using the Sigma-lognormal model.

**Formant Estimation**

To estimate the speech kinematics from the acoustic space in a non-invasive fashion, the formants are evaluated from the speech recorded with a microphone. This procedure is similar to the one used in handwriting, where the movements of the pencil tip are captured with a digitizing tablet.

Formants can be tracked using many methods proposed in the literature. In this paper, we use some of the methods implemented in the PRAAT software [43] to ensure experimental repeatability and to test the dependence of the proposed methodology on the formant estimation procedure.

Since there are no clear formants for unvoiced consonants, in fluent speech, they are usually co-articulated with a voiced sound [35], and so we assume that the missing formant information can be interpolated as a movement from the positions of the previous and posterior voiced phonemes.

**Formants to End Effector Kinematics**

A speech kinematics can be computed from its speech formants since the formant track is related to the movement in the tube resonance model and its velocity. Usually, the first two formants of the voice ($F_1$ and $F_2$) can give a spatial representation of the most frequent sounds and can be used to estimate the movement of the end effector needed to go from one sound to another. As can be seen in Fig. 4 (left), increments or decrements in the first or the second formant are related to changes in the pronounced sound. These changes can be represented as a trajectory drawn on an imaginary axis (Fig. 4 (right)). Since the proportion of the contribution of the first and second formants to the kinematic space is an ill-posed problem [39, 44–46], a transfer coefficient $c_i$ is added to the mapping equation. Hence, the conversion from the acoustic space to the kinematics space can be approximated by a linear transform such as

$$\begin{cases} \frac{\partial y(t)}{\partial t} = c_1 \frac{\partial F_1(t)}{\partial t} \\ \frac{\partial x(t)}{\partial t} = c_2 \frac{\partial F_2(t)}{\partial t} \end{cases} \quad (8)$$

where $F_1(t)$ and $F_2(t)$ are the tracks in the first and second formants, $c_i$ are the transfer coefficients, $x(t)$ and $y(t)$ are the trajectories along the two imaginary axes in the kinematic space, $\partial x(t)/\partial t$ denotes the derivative of the generic sequence $x(t)$, and $\partial y(t)/\partial t$ denotes the derivative of the generic sequence $y(t)$.

Once $x(t)$ and $y(t)$ are calculated from the formants, the approximate velocity $v_f(t)$ is estimated as

$$v_f(t) = \sqrt{\left(\frac{\partial x(t)}{\partial t}\right)^2 + \left(\frac{\partial y(t)}{\partial t}\right)^2} = \sqrt{\left(\frac{\partial (c_2 F_2))}{\partial t}\right)^2 + \left(\frac{\partial (c_1 F_1))}{\partial t}\right)^2} \quad (9)$$

The end effector reference center (RC) trajectory can thus be obtained by integrating Eq. 8, which leads to:

$$\begin{cases} y(t) = c_1 F_1(t) \\ x(t) = c_2 F_2(t) \end{cases} \quad (10)$$

We assume that the initial conditions, which are irrelevant for the Sigma-lognormal analysis, are equal to zero. Then, $x(t)$ and $y(t)$ refer to the spatiotemporal sequence that represents the end effector movement. Thus, $c_1$ and $c_2$ can be seen as the weights that map the formants $F_1$ and $F_2$ into their spatial representation. To allow evaluating the proportion between $F_1$ and $F_2$ that can give more information regarding the articulatory movement to which the kinematic theory of rapid human movements should be applied, in this work, we novelly calculate these weights using

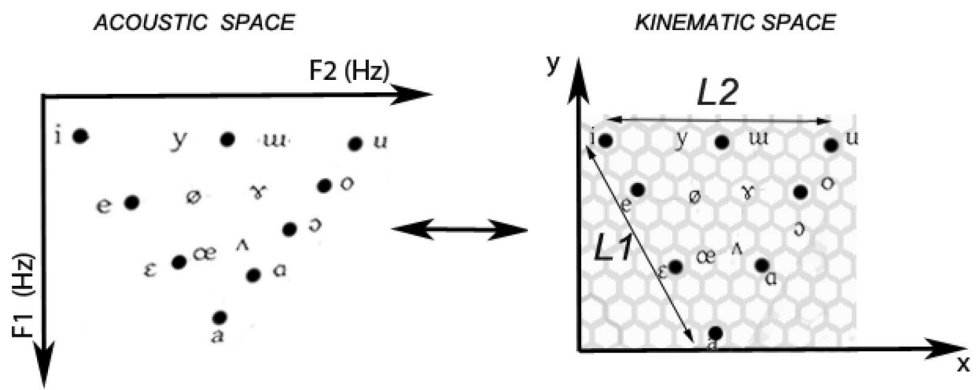

**Fig. 4** Space transformation. Left: acoustic space, right: kinematic space

$$\begin{cases} c_1 = k(1-\alpha) \\ c_2 = k\alpha \end{cases} \quad 0 \le \alpha \le 1 \tag{11}$$

where $k$ is the scale constant and $\alpha$ is the proportion parameter that defines the relative contribution of $F_1$ and $F_2$ to the kinematic space. It depends on the shape of the vocal tract.

To calculate $\alpha$ and $k$, we assume that the acoustic space could be transformed into a hexagonal kinematic space (Fig. 4, right). Based on previous handwriting synthesis studies [20], and inspired by the hexagonal grid cell distribution proposed by Moser et al. [27, 28], the vowel triangle limits are fitted with an equilateral triangle. In this work, we hypothesize that $\alpha$ can be approximated as the value that keeps $L_1 = L_2$, with $L_1$ being the distance between the /i/ position and /a/, and $L_2$, the distance between the /i/ and /u/ (Fig. 4, right).

To this end, as the external vowels of the kinematic space are usually /a/, /i/, and /u/ (see Fig. 4), we define $F_{1a}$, the first formant of the vowel /a/, $F_{1i}$ and $F_{2i}$, the first and second formants of the vowel /i/, respectively, and $F_{2u}$, the second formant of the vowel /u/.

The height of the triangle can be calculated as

$$h = c_1(F_{1a} - F_{1i}) \tag{12}$$

Considering the triangle as an equilateral triangle, it means that

$$L_1 = \frac{2}{\sqrt{3}} h = \frac{2}{\sqrt{3}} c_1(F_{1a} - F_{1i}) \tag{13}$$

And

$$L_2 = c_2(F_{2i} - F_{2u}) \tag{14}$$

As $L_1 = L_2$,

$$\frac{2}{\sqrt{3}} c_1(F_{1a} - F_{1i}) = c_2(F_{2i} - F_{2u}) \tag{15}$$

Replacing $c_1$ and $c_2$ by their values:

$$\frac{2}{\sqrt{3}} k(1-\alpha)(F_{1a} - F_{1i}) = k\alpha(F_{2i} - F_{2u}) \tag{16}$$

Obtaining $\alpha$ (the proportion between $F_1$ and $F_2$):

$$\alpha = \frac{(F_{1a} - F_{1i})}{(F_{1a} - F_{1i}) + \frac{\sqrt{3}}{2}(F_{2i} - F_{2u})} \tag{17}$$

It should be noted that both the value of $\alpha$ and the formant values are speaker-dependent. Table 1 shows the $\alpha$ values obtained with Eq. 17 using the formant values given by Hillenbrand et al. in [47] (English vowels) and by Pätzold et al. [48] (German vowels). We can see that the values range from 0.25 to 0.33.

**Table 1** Value estimated from different previous works

| Reference | Gender | $\alpha_{mean}$ | $\alpha_{min}$ | $\alpha_{max}$ |
| --- | --- | --- | --- | --- |
| Hillenbrand et al. in (47) (English) | Male | 0.27 | 0.25 | 0.33 |
|  | Female | 0.258 | 0.25 | 0.33 |
| Pätzold (48) (German) | Male | 0.28 | 0.26 | 0.30 |
|  | Female | 0.29 | 0.25 | 0.31 |

The constant $k$ is a scale factor that converts the estimated values of $x(t)$ and $y(t)$ to centimeters. Unlike with the proportion parameter $\alpha$, this constant is not necessary for the Sigma-lognormal model. However, a reasonable value of $k$ facilitates understanding of the model.

To find this reasonable value, we can use the already known relationship between $L_2$ and $k$ given by:

$$L_2 = c_2(F_{2i} - F_{2u}) = k\alpha(F_{2i} - F_{2u}) \tag{18}$$

$k$ is thus obtained as

$$k = \frac{L_2}{\alpha(F_{2i} - F_{2u})} \tag{19}$$

To calculate the $k$ value associated with a real movement, the $L_2$ value can be obtained from the results presented by Whitfield et al. [49], where the movement needed to utter the sentence "It's time to shop for two new suits" was measured in 20 subjects with sensors. We take the values obtained with the tongue front marker (TF) in mm, the mean of the range of $F_2$, and the parameter $\alpha$ rounded to 0.3. This leads to a $k$ factor of about 0.04 mm/Hz, which keeps the peak velocities similar to the ones presented in [50].

### Sigma-Lognormal Analysis

Once the trajectory has been estimated, it is modeled with the kinematic theory of rapid human movements through the Sigma-lognormal model, as is explained in "Overview of the Sigma-Lognormal Model". The kinematic theory is applied in an attempt to model the speech kinematics as a synchronized summation of simple overlapped movements, inspired by how the brain issues time-spaced commands to the articulatory organs. As such, speech is modeled as a global movement instead of a single muscle or group of muscles modeled independently.

The hypothesis underlining the application of this model to speech posits that a lognormal in speech has a similar meaning as in handwriting, a primitive that has been widely tried and tested. Therefore, in the case of speech, the number of lognormals would be related to the number of simple articulatory movements for a natural and healthy speech. Hence, the number of lognormals should be related to the number of speech sounds uttered and their timing. Obviously, it is expected that a neuromotor dysfunction will affect the number, shape, and time of occurrence of the lognormals, as is the case in handwriting [19]. These neuromotor dysfunctions can be due to normal aging or neurodegenerative diseases. In the special case of laryngeal pathologies, which affect only the closing of the glottis and voice source, they should not affect the timing parameters and the lognormal shape for subjects of the same age, but they could result in more simple movements due to the effort needed to talk and to the pauses in the pronunciation of a sentence.

Beyond the sequence of lognormal parameters $P = \{D_j, t_{oj}, \mu_j, \sigma_j, \Theta_{ej}, \Theta_{sj}, VTP_{j-1}\}_{j=1}^{NbLog}$, it makes sense to define and use supplementary parameters related to the timing intervals between lognormals and lognormal shapes. Such parameters can help improve our understanding of some diseases. Examples of these parameters include:

- $\overline{\Delta t_o}$: the mean of the time between successive lognormals, that is, the mean of the time difference between the current lognormal and the previous one:

$$\overline{\Delta t_o} = \frac{\sum_{j=2}^{NbLog}(t_{oj} - t_{o(j-1)})}{NbLog} \quad (20)$$

- $\overline{V_p}$: the average of the maximum velocity of the *Nblog* lognormals:

$$\overline{V_p} = \frac{\sum_{j=2}^{NbLog} \max(\vec{v}_j(t))}{NbLog} \quad (21)$$

- $\overline{\mu}$: the mean of the log time delay:

$$\overline{\mu} = \frac{\sum_{j=1}^{NbLog} \mu_j}{NbLog} \quad (22)$$

- $\overline{\sigma}$: the mean of the lognormal response time:

$$\overline{\sigma} = \frac{\sum_{j=1}^{NbLog} \sigma_j}{NbLog} \quad (23)$$

- $\overline{D}$: the mean of the lognormal distance covered in the kinematic space:

$$\overline{D} = \frac{\sum_{j=1}^{NbLog} |D_j|}{NbLog} \quad (24)$$

## Evaluation, Results, and Discussion

The evaluation of the model is aimed at answering the following three questions:

1. What is the meaning of each lognormal in speech?
2. Which range of $\alpha$ (the proportion between $F_1$ and $F_2$) is adequate to apply the Sigma-lognormal model?
3. Do the speech lognormal parameters model aging phenomena in speech?

### Databases

In handwriting, a lognormal expresses a primitive movement, related to a simple stroke. If a lognormal in speech retains a similar meaning, strokes should be associated with simple speech movements that are linked to the movements needed to pronounce a speech sound. To check this hypothesis, we used the VTR-TIMIT database [51, 52]. The advantage of this database is that the formants it contains have been manually annotated and the phonemes labeled, providing the background that allows correlating lognormals to phonemes.

The VTR-TIMIT [51] database is composed of 538 English utterances from the TIMIT corpus [52], with phonetically compact sentences (SX) and phonetically diverse sentences (SI). The VTR-TIMIT database is labeled by phonemes. In this experiment, we use the complete dataset of 197 speakers and 538 utterances in total. The database is balanced in terms of speakers, dialects, gender, and phonemes [51].

Furthermore, as the Sigma-lognormal model links lognormals to the impulse response of a neuromotor system, it is assumed that only neurodegenerative or neuromotor diseases will affect the lognormal parameters in fluent speech. To assess this premise, we used the Saarbruecken Voice Database [53]. This database contains healthy speakers as well as speakers with laryngeal pathologies. The database is labeled with the speaker's age and the kinds of pathologies they have.

The Saarbruecken Voice Database [53] is a collection of German speech recordings from more than 2000 speakers. The sentence recorded is "Guten Morgen, wie geht es Ihnen?" ("Good morning, how are you?"). For our experiments, we divided this database into three groups:

- Young speakers' group, which encompassed speakers aged between 20 and 30. It included both healthy speakers and speakers with laryngeal pathologies. There was a total of 609 speakers, with 236 males and 373 females.

- Middle speakers' group, which encompassed speakers aged between 40 and 50, containing healthy speakers and speakers and with laryngeal pathologies. There were a total of 352 speakers: 177 males and 175 females.
- Older speakers' group, which included all speakers aged 60 to 80 years old. All in all, there were 466 speakers in this group: 262 males and 204 females.

All the recordings were made in a controlled environment at a sampling frequency of 50 kHz and a 16-bit resolution. The recordings contained 71 different laryngeal pathologies, including some organic and functional members.

The term *laryngeal pathologies* (LP) comprises a wide range of disorders, the most frequent ones being organic, and affecting the morphology of the excitation organs and producing irregular vibration patterns [54]. Some examples of these disorders are polyps, nodules, edemas, and carcinomas. The phonation in these cases is characterized by noisy bands in the spectrogram, instability in the vibration frequency of the vocal cords, irregular airflow, and the presence of turbulent noise.

### Experiment 1: Meaning of Lognormal in Speech and $\alpha$ Empirical Estimation

To assess the meaning of a lognormal in speech, the first experiment aimed to study the relationship between lognormals and phonemes. Additionally, as the velocity is a function of $\alpha$ (the proportion between formants), the optimum values of this constant are estimated in this experiment to be compared with the theoretical estimation in "Formants to End Effector Kinematics."

For this assessment, we employed the publicly available VTR-TIMIT database of continuous speech, which is labeled by phonemes, thus providing the number of phonemes ($N_p$) in each sentence. All the sentences of this database were analyzed by ScriptStudio [31] and decomposed into a sequence of lognormals. The number and timing of lognormals were compared with the phonemic labels of the database. The velocity was obtained from the formant track provided by the dataset.

An example of such an analysis is shown in Fig. 5. It corresponds to an excerpt ("Their records") from the sentence "How permanent are their records?" in English. In this figure, we can observe the speech waveform, the spectrogram with its formant track, and the lognormal decomposition of the velocity. In Fig. 5, it can be seen that there are almost as many phonemes as there are lognormals. Besides, the lognormals are temporally ahead of the phoneme as the movement between two phonemes precedes the sound. This is shown in Fig. 6, which is zoomed in Fig. 5. Further, we can observe how the velocity peak usually appears alongside the phoneme transition, since a fast change in the resonance cavities is required to pronounce the next sound. As well, we can see that when the duration of the phonemes is long or the articulation of the phoneme requires the pronunciation of more than one simple movement, more than one lognormal appears, as is the case between 1.1 and 1.2 s in Fig. 5.

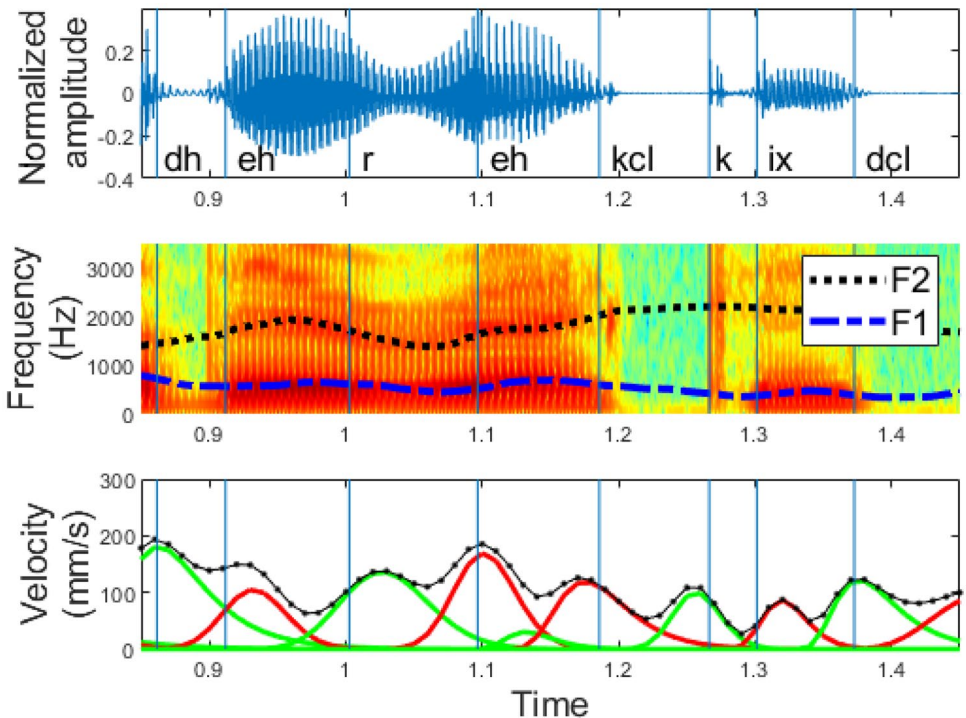

**Fig. 5** Relationship between phonemes and lognormals. Figure at the top: speech signals ("their records") segmented by phonemes, figure in the middle: spectrogram, and figure at the bottom: lognormal decomposition (color changes for even and odd lognormals), velocity profile (dotted line), and TIMIT phoneme segmentation (blue bars)

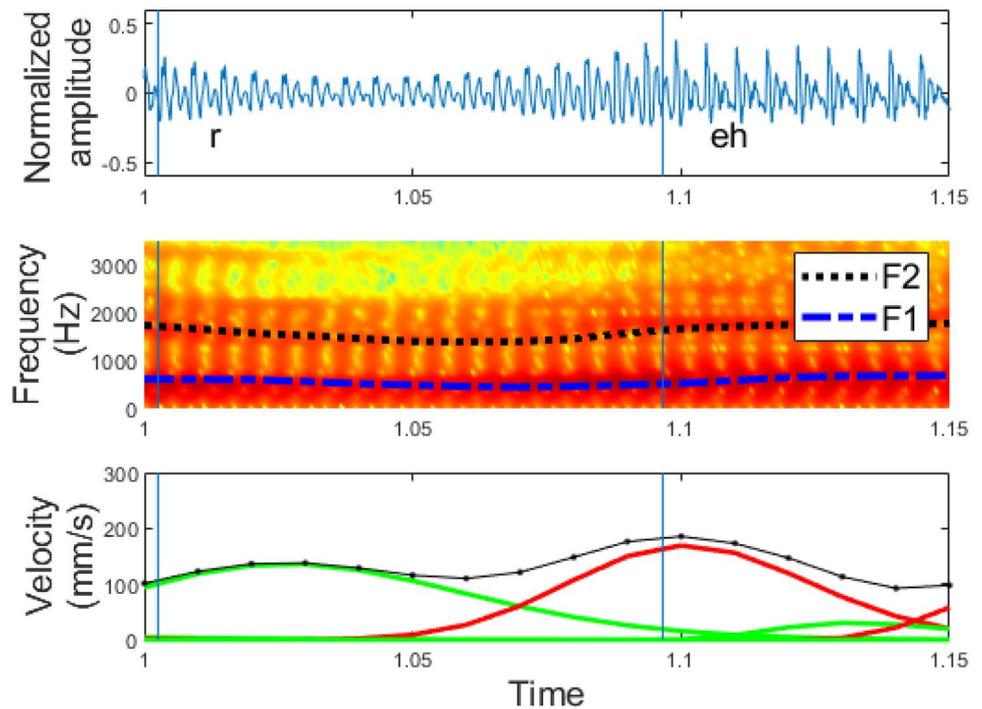

**Fig. 6** Room-in of Fig. 5 to show the exact correspondence between phoneme and lognormal

To illustrate how the correspondence between the phonemes and lognormals is obtained from a sentence, a study was carried, looking at one phoneme after the other. For each phoneme, four possibilities were considered:

1. True positive (*TP*): A lognormal of the sentence that overlaps the phoneme is assigned to it. In this case, $TP_i = 1$, with i being the index of the phoneme.
2. False positive (*FP*): Other lognormals of the sentence that overlap the phoneme in study. In this case, $FP_i$ is set to the number of lognormals that overlap the phoneme minus 1.
3. False negative (*FN*): If no one lognormal overlaps the phoneme, $FN_i$ is set to 1.
4. True negative (*TN*): The set of lognormals that belong to the sentence do not overlap the phoneme. In this case, $TN_i$ is set to the number of lognormals that do not overlap the phoneme.

Note that $TP_i + FP_i + TN_i$ is equal to the number of lognormals of the sentence. The bounds of the lognormals are considered at 5% of its peak value. The measurements of the matching between the phonemes of the sentence and the lognormals obtained with the sentence are given in terms of the true positive rate and true negative rate of the sentence and are calculated as $TPR_s = \left(\sum_{i=1}^{N_p} TP_i\right)/N_p$ and $TNR_s = \left(\sum_{i=1}^{N_p} TN_i\right)/(NbLog - 1)$, respectively. The *TPR* and *TNR* of the VTR-TIMIT dataset are obtained by averaging the *TPRs* and *TNRs* of all the sentences in it.

Figure 7 shows *TPR* and *TNR* curves per gender and the mean value of both as a function of $\alpha$. Although this $\alpha$ value used to work out the velocity from the formant track Eqs. 9-11 was obtained theoretically in Eq. 17, it can be empirically validated to obtain the *TPR* and *TNR* for different $\alpha$ values.

Moreover, to see the correlation between the velocity peak occurrence ($t_v$) and the phoneme transition ($t_p$), the relationship between them, as seen in Fig. 8, is obtained through the error rate ($\varepsilon_t$) as:

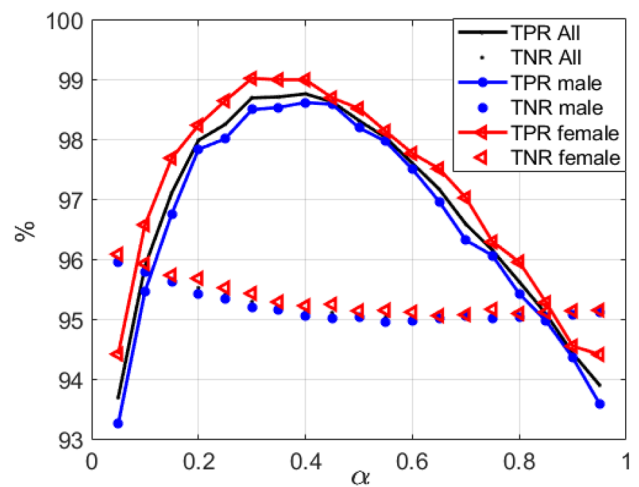

**Fig. 7** TPR and TNR curves across the VTR-TIMIT database as a function of F1-F2 proportion value $\alpha$. The value of the lognormal is greater than the 5% of its peak value

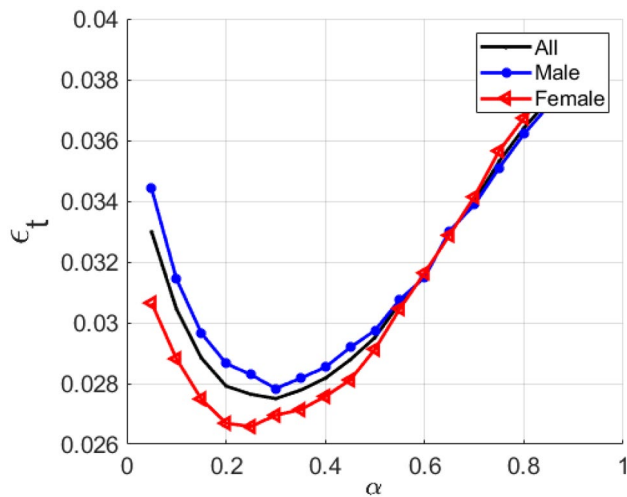

**Fig. 8** Error rate across the VTR-TIMIT database as a function of $F_1$-$F_2$ proportion value $\alpha$

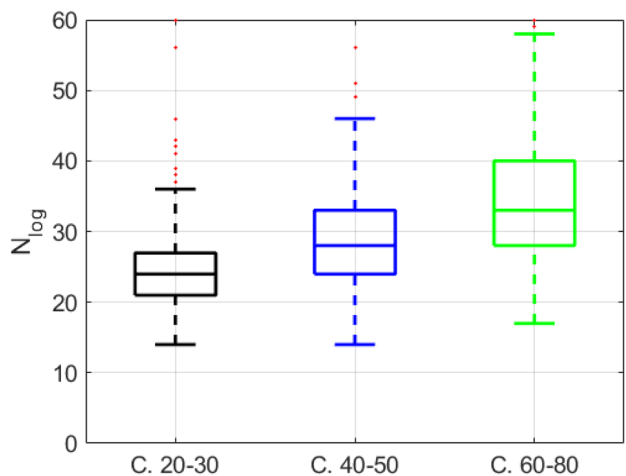

**Fig. 9** Comparison of the number of lognormals obtained with the young, middle, and older groups of speakers obtained from the Saarbruecken Voice Database, obtaining the formants with "To formants (sl)" method

$$\varepsilon_t = \sqrt{\frac{1}{Np} \sum_{i=1}^{Np} (t_{vi} - t_{pi})^2} \qquad (25)$$

For the experiments, although the value of $k$ (see Eq. 19) does not affect the velocity profile shape or the result, it is approximated to 0.04 to keep a velocity peak close to 200 mm/s as measured by the sensors in [50].

We can see in Fig. 7 that the *TPR* curves get the maximum values of $\alpha$ around to 0.35. Further, as seen in Fig. 8, the $\varepsilon_t$ gets minimum values for $\alpha$ lying between 0.2 and 0.4, which means that the lognormal peak is closest to the phoneme in healthy adults. These results show that for both males and females, this procedure is effective and not overly sensitive to the value of $\alpha$ in the $0.2 \leq \alpha \leq 0.4$ range, which is similar to the value proposed in "Formants to End Effector Kinematics.". Note that to pronounce some phonemes, more than one simple movement is required, and each subject could need a different $\alpha$ value.

### Experiment 2: Speech Lognormals, Aging, and Laryngeal pathologies

As speakers get older, their neuromotor systems deteriorate and movements require additional effort and become slower. In handwriting, this implies additional short strokes and slow handwriting. The same should apply to speech: a greater number of short movements or small lognormals and more time between these lognormals than in young speakers.

In this context, to gain insights into the meaning of a lognormal representation in speech, the second experiment compared the lognormals detected in young and older speakers, including subjects with laryngeal pathologies.

The experiment was run with the Saarbruecken Voice Database, which labels recorded sentences with the age of the speakers and allows comparisons between the results obtained with the groups of young and older speakers. In the cases where result shows a significant difference (*NbLog*, $\overline{\Delta t_o}, \overline{\mu}$, SNR) the experiments were repeated in order to evaluate the evolution of the parameters along three age groups (young, middle, and older) (Table 5). Gender is omitted in the analyses that follow since the experiments in "Experiment 1: Meaning of Lognormal in Speech and Empirical Estimation" show similar results for males and females.

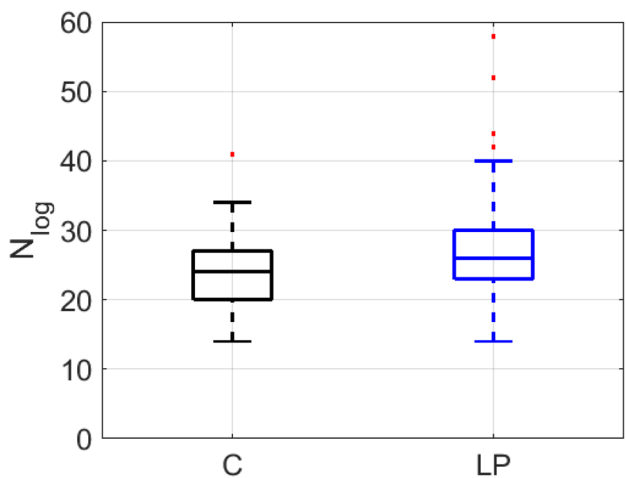

**Fig. 10** Comparison of the number of lognormals obtained from healthy young speakers (C) and speakers with laryngeal pathology (LP)

**Table 2** Averaged and standard deviation value of the lognormal parameters for young and older groups (parameter with statistically significant difference in italic)

| Parameter | Formant estimation method | | | | | |
|---|---|---|---|---|---|---|
| | "To formant (keep all)" | | | "To formant (sl)" | | |
| | Young (20–30) | Older (60–80) | *p* value | Young (20–30) | Older (60–80) | *p* value |
| $\overline{\Delta t_o}$ | *0.066 ± 0.008* | *0.068 ± 0.008* | *< 0.001* | *0.067 ± 0.007* | *0.07 ± 0.007* | *< 0.001* |
| *NbLog* | *25 ± 6.03* | *33 ± 9.48* | *< 0.001* | *24 ± 6.2* | *34 ± 9.4* | *< 0.001* |
| $\overline{\sigma}$ | 0.12 ± 0.04 | 0.12 ± 0.03 | 0.22 | 0.14 ± 0.05 | 0.13 ± 0.04 | 0.0623 |
| $\overline{\mu}$ | − 1.55 ± 0.16 | − 1.52 ± 0.14 | 0.16 | *− 1.6 ± 0.17* | *− 1.56 ± 0.14* | *< 0.001* |
| $\overline{V}_p$ | 23.17 ± 4.3 | 23.84 ± 5 | 0.06 | 14.2 ± 2.19 | 13.87 ± 2.1 | 0.055 |
| $\overline{D}$ | 1.22 ± 0.22 | 1.24 ± 0.24 | 0.23 | 0.8 ± 0.12 | 0.78 ± 0.04 | 0.14 |
| *SNR* | *19.95 ± 3.4* | *19.23 ± 3.7* | *< 0.001* | *18.27 ± 3.58* | *17.55 ± 3.54* | *0.002* |

Moreover, gender is reasonably balanced in the database, and the effect of age and gender cannot be confounded [19].

As the Saarbruecken Voice Database does not provide formant tracks, these were obtained with the following two formant estimation methods (available in the Praat software package [43]):

- "From speech to formant (sl)": This algorithm is based on the implementation of the Split Levinson algorithm proposed by Willems [55]. It always finds the requested number of formants in every frame, even if they do not exist.

- "From speech to formant (keep all)": In this case, Praat applies a Gaussian-like window and computes the formant from the LPC spectrum obtained through the Burg algorithm [56, 57].

The following settings were used in the Praat software for both methods to determine the first two formants in all the sentences of the Saarbruecken Voice Database: time step of 0.01 s, maximum number of formants of 5, and window length of 0.025 s.

To calculate the speech kinematics, based on the previous result, the parameter $\alpha$ was set to 0.3 and $k$ to 0.04. The speech trajectory was processed with ScriptStudio® [30] to decompose the speech kinematics into lognormals.

The results are graphically shown in Figs. 9 and 10, and numerically in Tables 2-5. These tables also include the averaged values and the standard deviation of all the lognormal parameters, along with a one-way ANOVA (analysis of variance) [58]. Multiple comparison tests with Bonferroni correction are used when three classes (young, middle, and older) are analyzed. In this type of analysis, two groups are considered as statistically different if the residual *p* value is below 0.05 and statistically similar if the *p* value is above 0.05 [58].

The findings can be summarized as follows:

1. $\overline{\Delta t_o}$ is sensitive to the speaker's age. While there is a significant difference between the young versus the older speakers (*p* value <0.001), there is no significant statistical variation between healthy speakers and speakers with laryngeal pathologies for this parameter (Tables 3-5). This means that the time between commands increases with age [19], due to the increase in $\Delta t_0$, but not with laryngeal pathologies. This is consistent with the well-known fact regarding slower reaction times in these conditions.

2. The number of lognormals (*NbLog*) is greater for the group of older speakers than for the group of young

**Table 3** Averaged value of the lognormal parameters for young healthy speakers and speakers with laryngeal pathologies (parameter with statistically significant difference in italic)

| Parameter | Formant estimation method | | | | | |
|---|---|---|---|---|---|---|
| | "To formant (keep all)" | | | "To formant (sl)" | | |
| | Control (20–30) | LP (20–30) | *p* value | Control (20–30) | LP (20–30) | *p* value |
| $\overline{\Delta t_o}$ | 0.067 ± 0.008 | 0.066 ± 0.008 | 0.06 | 0.067 ± 0.0077 | 0.067 ± 0.007 | 0.97 |
| *NbLog* | *24 ± 4.5* | *26 ± 7.6* | *< 0.001* | *23 ± 4.7* | *26 ± 8* | *< 0.001* |
| $\overline{\sigma}$ | 0.12 ± 0.05 | 0.118 ± 0.05 | 0.32 | 0.15 ± 0.04 | 0.15 ± 0.05 | 0.71 |
| $\overline{\mu}$ | − 1.55 ± 0.18 | − 1.54 ± 0.14 | 0.46 | − 1.60 ± 0.17 | − 1.60 ± 0.17 | 0.5 |
| $\overline{V}_p$ | 23.68 ± 4.81 | 23.38 ± 4.04 | 0.74 | 14.2 ± 2.18 | 14.14 ± 1.93 | 0.98 |
| $\overline{D}$ | 1.25 ± 0.25 | 1.22 ± 0.2 | 0.33 | 0.79 ± 0.12 | 0.79 ± 0.1 | 0.84 |
| *SNR* | *20 ± 3.2* | *19.5 ± 3.5* | *0.01* | *18.98 ± 3.56* | *17.62 ± 3.65* | *0.04* |

**Table 4** Averaged value of the lognormal parameters for older healthy speakers and speakers with laryngeal pathologies

| Parameter | Formant estimation method | | | | | |
|---|---|---|---|---|---|---|
| | "To formant (keep all)" | | | "To formant (sl)" | | |
| | Control (60–80) | LP (60–80) | $p$ value | Control (60–80) | LP (60–80) | $p$ value |
| $\overline{\Delta t_o}$ | 0.069 ± 0.009 | 0.068 ± 0.008 | 0.64 | 0.068 ± 0.007 | 0.069 ± 0.007 | 0.50 |
| $NbLog$ | 30 ± 8.8 | 33 ± 8.5 | 0.05 | 30 ± 8.2 | 33 ± 9.5 | 0.13 |
| $\overline{\sigma}$ | 0.10 ± 0.04 | 0.12 ± 0.04 | 0.17 | 0.13 ± 0.04 | 0.14 ± 0.04 | 0.39 |
| $\overline{\mu}$ | −1.47 ± 0.17 | −1.52 ± 0.14 | 0.18 | −1.58 ± 0.14 | −1.56 ± 0.14 | 0.53 |
| $\overline{V}_p$ | 23.81 ± 4.5 | 23.82 ± 5 | 0.34 | 14.3 ± 1.95 | 13.86 ± 2.1 | 0.40 |
| $\overline{D}$ | 1.20 ± 0.24 | 1.25 ± 0.25 | 0.39 | 0.76 ± 0.1 | 0.78 ± 0.1 | 0.71 |
| $SNR$ | 18.58 ± 3.1 | 19.3 ± 3.7 | 0.52 | 18.14 ± 3.7 | 17.74 ± 3.5 | 0.88 |

speakers (Tables 2 and 5; Fig. 9). The $p$ value is lower than 0.001 with both formant estimation algorithms. This is consistent with the results observed in handwriting, where the kinematic theory was used to evaluate aging. The results might suggest that the deterioration of motor control with aging is associated with the development of compensatory strategies such as emitting more motor commands to generate an adequate movement for a given task [19]. A significant difference is also observed between young healthy and LP speakers in the number of lognormals Fig. 10, Table 3). This type of disease should, therefore, influence the number of lognormals due to increases in the number of simple movements following necessary pauses and silences in a sentence.

3. The SNR parameter decreases and the number of lognormals ($NbLog$) increases in both older people and the LP group in young people (Table 3), with the difference being more significant with age than with laryngeal disease.
4. The $\overline{\mu}$ parameter increases with age, indicating that the impulse response of the system is slower in the case of older speakers. This difference is only appreciated with the "from speech to formant (sl)" method (Table 2), and this could be because this formant extraction method always gives the requested number of formants in every frame, allowing the best interpolation of the complete movement in the case of consonants.
5. Regarding the parameters $\overline{\sigma}$, $\overline{V}_p$, and $\overline{D}$, in Tables 2-4, no significant difference can be seen between the two age groups.
6. If we compare the three age groups (Table 5) only with the $NbLog$ parameter, significant differences are found between the three classes (Fig. 9).

## Discussion

The results show how the Sigma-lognormal model can be applied to model neuromuscular aging in speech. When the speech is modeled, each of the lognormals obtained reflects a group of commands and their end muscular response shapes. Neurological diseases, learning processes, or aging can affect this command sequence, changing the proportion of final movements, the speech rate, or the muscular response shape, which is consistent with the lognormality principle [19].

In the above results, the parameters related to the time between commands ($\overline{\Delta t_o}$) and the delay in the muscular response ($\overline{\mu}$) are longer in older people, as the movements become slower with age. Moreover, the experiments show a clear relationship between the number of simple movements found by the model and the number of pronounced phonemes. This relationship is conditional on the proportion between the first and second formants used to estimate the trajectory, as we tested with the experiments. Also, the method used to detect the formants can affect the parameters obtained, providing the Sigma-lognormal method with information on how the formant extractor is able to follow muscular movements.

**Table 5** Averaged and STD values of the lognormal parameters for young, middle, and older speakers with laryngeal pathologies ("To formant (sl)")

| Parameter | Young (40–30) | Middle (40–50) | Older (60–80) | $p$ value Y vs M | $p$ value Y vs O | $p$ value M vs O |
|---|---|---|---|---|---|---|
| $\overline{\Delta t_o}$ | 0.067 ± 0.007 | 0.069 ± 0.007 | 0.07 ± 0.007 | 0.02 | < 0.001 | 1 |
| $NbLog$ | 24 ± 6.2 | 39.31 ± 7.8 | 34 ± 9.4 | < 0.001 | < 0.001 | < 0.001 |
| $\overline{\mu}$ | −1.6 ± 0.17 | −1.59 ± 0.15 | −1.56 ± 0.14 | 1 | < 0.001 | 0.007 |
| $SNR$ | 18.27 ± 3.58 | 18.17 ± 3.46 | 17.55 ± 3.54 | 1 | < 0.001 | 0.05 |

The model was tested with two different languages (English and German) and seems to be language-independent, as has also been observed when the lognormal model is applied in handwriting [59].

# Conclusions

A Sigma-lognormal representation for modeling speech kinematics has been presented. The speech kinematics is estimated from the formant tracks and decomposed into simple lognormal movements by applying the kinematic theory of rapid human movements. Moreover, besides the Sigma-lognormal parameters, a set of derived parameters is proposed to describe the timing and the neuromotor impulse response.

The experiments conducted illustrate the lognormal meaning in speech and indicate the adequate relation between first and second formants in order to get the kinematic information. The first experiment shows the link between a lognormal and a transition between phonemes, where the number of the lognormals is similar to that of phonemes. In this experiment, that the optimum proportion between the first and second formants was also verified. The second experiment links the lognormal to the generation of each end effector movement, showing that the parameter $\overline{\Delta t_o}$, as in handwriting, increases significantly from young to older speakers, and that it is independent of dysfunction, such as problems in the larynx or glottis closure. This allows modeling aging in speech production as a delay between commands and the end effector responses.

The results show that it is possible to model speech with the kinematic theory, which provides biological information about the simple movements involved in speech.

As future lines of research, the model could be applied to speech synthesis, speech recognition, speech rehabilitation, as well as to the design of systems to help in the screening and monitoring of some neurodegenerative diseases. The model could also permit the use of features similar to those obtained from studying other human movements, such as handwriting. Moreover, investigating the use of more formants to estimate speech kinematics is an unresolved issue that is yet to be addressed.

**Funding** This study was funded by the Spanish government's MIMECO TEC2016-77791 research project and European Union FEDER program/funds, Teca-Park/MonParLoc FGCSIC CENIE-0348_CIE_6_E (InterReg Programme) to Pedro Gomez-Vilda, the NSERC-Canada Grant RGPIN-2015-06409 to R. Plamondon.

C. Carmona-Duarte was supported by a Juan de la Cierva contract (IJCI-2016-27682), Viera y Clavijo grant from ULPGC and the "José Castillejo" mobility grant from the Spanish government CAS18/00315.

## Compliance with Ethical Standards

**Conflict of Interest** The authors declare that they have no conflict of interest.

**Ethical Standards** This article does not contain any studies with human participants performed by any of the authors.

**Informed Consent** Informed consent was obtained from all individual participants included in the study.

**Research Involving Human and Animal Rights** This paper does not contain any studies with animals performed by any of the authors.

## References

1. Guenther FH. Speech sound acquisition, coarticulation, and rate effects in a neural network model of speech production. Psychol Rev. 1995;102(3):594–621.
2. Parrell B, Lammert AC, Ciccarelli G, Quatieri TF. Current models of speech motor control: a control-theoretic overview of architectures and properties. J Acoust Soc Am. 2019;145(3):1456–81.
3. Perrier P, Ma L, Payan Y. Modeling the production of VCV sequences via the inversion of a biomechanical model of the tongue. 9th Eur Conf Speech Commun Technol. 2019;1041–4.
4. Patri JF, Diard J, Perrier P. Optimal speech motor control and token-to-token variability: a Bayesian modeling approach. Biol Cybern. 2015;109(6):611–26.
5. Kröger BJ, Kannampuzha J, Neuschaefer-Rube C. Towards a neurocomputational model of speech production and perception. Speech Commun. 2009;51(9):793–809.
6. Tourville JA, Guenther FH. The DIVA model: a neural theory of speech acquisition and production. Lang Cogn Process. 2011;26(7):952–81.
7. Saltzman EL, Munhall KG. A dynamical approach to gestural patterning in speech production. Ecol Psychol. 1989;1(4):333–82.
8. Houde JF, Nagarajan SS. Speech production as state feedback control. Front Hum Neurosci. 2011;5(October):1–14.
9. Parrell B, Ramanarayanan V, Nagarajan S, Houde J. The FACTS model of speech motor control: fusing state estimation and task-based control. PLoS Comput Biol [Internet]. 2019;15(9):1–26. Available from: https://doi.org/10.1371/journal.pcbi.1007321.
10. Plamondon R, O'Reilly C, Galbally J, Almaksour A, Anquetil É. Recent developments in the study of rapid human movements with the kinematic theory: applications to handwriting and signature synthesis. Pattern Recognit Lett. 2014;35(1):225–35.

11. Plamondon R. A kinematic theory of rapid human movements. Part I: Movement representation and generation. Biol Cybern [Internet]. 1995;72(4): 295–307. Available from: https://www.ncbi.nlm.nih.gov/pubmed/7748959.
12. Plamondon R. A kinematic theory of rapid human movements. Part II: Movement time and control Biol Cybern. 1995;72(4):309–20.
13. Plamondon R. A kinematic theory of rapid human movements. Part III: Kinematic Outcomes Biol Cybern. 1998;78(2):133–45.
14. Plamondon R, Pirlo G, Anquetil É, Rémi C, Teulings HL, Nakagawa M. Personal digital bodyguards for e-security, e-learning and e-health: a prospective survey. Pattern Recognit. 2018;81:633–59.
15. Leiva LA, Martín-Albo D, Plamondon R. The kinematic theory produces gestures. Human-like Stroke Interact Comput. 2017;29(4):552–65.
16. Lebel K, Nguyen H, Duval C, Plamondon R, Boissy P. Capturing the cranio-caudal signature of a turn with inertial measurement systems: methods, parameters robustness and reliability. Front Bioeng Biotechnol [Internet]. 2017;5:1–13. Available from: http://journal.frontiersin.org/article/10.3389/fbioe.2017.00051/full.
17. Martín-Albo D, Leiva LA, Huang J, Plamondon R. Strokes of insight: user intent detection and kinematic compression of mouse cursor trails. Inf Process Manag. 2016;52(6):989–1003.
18. Nadeau A, Lungu O, Duchesne C, Robillard MÈ, Bore A, Bobeuf F, et al. A 12-Week cycling training regimen improves gait and executive functions concomitantly in people with parkinson's disease. Front Hum Neurosci [Internet]. 2017;10:1–10. Available from: http://journal.frontiersin.org/article/10.3389/fnhum.2016.00690/full.
19. Plamondon R, O'Reilly C, Rémi C, Duval T. The lognormal handwriter: learning, performing, and declining. Front Psychol. 2013;4:1–14.
20. Carmona-Duarte C, Ferrer MA, Parziale A, Marcelli A. Temporal evolution in synthetic handwriting. Pattern Recognit 2017;68.
21. Ferrer MA, Diaz M, Carmona C, Morales A. A behavioral handwriting model for static and dynamic signature synthesis. IEEE Trans Pattern Anal Mach Intell [Internet]. 2016;8828(c): 1. Available from: http://ieeexplore.ieee.org/document/7494603/.
22. Woch A, Plamondon R. Using the framework of the kinematic theory for the definition of a movement primitive. Mot Control. 2004;8(4):547–57.
23. Carmona-Duarte C, Góme-Vilda P, Ferrer MA, Plamondon R, Londral A. Study of several parameters for the detection of amyotrophic lateral sclerosis from articulatory movement. Loquens. 2017;4(January):1–5.
24. Carmona-Duarte C, Ferrer M, Gómez-Vilda P, Gemmert AWA Van. Plamondon R. A common framework to evaluate Parkinson's disease in voice and handwriting. In: ICPRAI 2018 - International Conference on Pattern Recognition and Artificial Intelligence. 2018.
25. Carmona-Duarte C, Plamondon R, Gómez-Vilda P, Ferrer MA, Alonso JB, Londral ARM. Application of the lognormal model to the vocal tract movement to detect neurological diseases in voice. In: Chen YW, Tanaka S, Howlett RJL, editors. Innovation in Medicine and Healthcare 2016 Smart Innovation, Systems and Technologies. Switzerland: Springer; 2016. p. 25–35.
26. Carmona-Duarte C, Alonso JB, Diaz M, Ferrer MA, Gómez-Vilda P, Plamondon R, et al. Kinematic modeling of diphthong articulation. In: Esposito A, Faundez-Zanuy M, Esposito AM, Cordasco G, Drugman T, Solé-Casals J, et al., editors. Recent Advances in Nonlinear Speech Processing. Cham: Springer; 2016. p. 53–60.
27. Hafting T, Fyhn M, Molden S, Moser M, Moser EI. Microstructure of a spatial map in the entorhinal cortex. Nature. 2005;436(7052):801–6.
28. Moser EI, Moser MB, Roudi Y. Network mechanisms of grid cells. Philos Trans R Soc B Biol Sci. 2014;369:1635.
29. Tremblay P, Sato M, Deschamps I. Age differences in the motor control of speech: an fMRI study of healthy aging. Hum Brain Mapp. 2017;38(5):2751–71.
30. O'Reilly C, Plamondon R. Development of a sigma-lognormal representation for on-line signatures. Pattern Recognit [Internet]. 2009;42(12):12:3324–37. Available from: https://doi.org/10.1016/j.patcog.2008.10.017.
31. Djioua M, Plamondon R. A new algorithm and system for the characterization of handwriting strokes with delta-lognormal parameters. IEEE Trans Pattern Anal Mach Intell. 2009;31(11):2060–72.
32. Ferrer MA, Diaz M, Carmona-Duarte C, Plamondon R. iDeLog: Iterative dual spatial and kinematic extraction of sigma-lognormal parameters. IEEE Trans Pattern Anal Mach Intell. 2018;PP(c):1.
33. Plamondon R, Feng C, Woch A. A kinematic theory of rapid human movement. Part IV: A formal mathematical proof and new insights. Biol Cybern 2003;89(2):126–38.
34. Marcelli A, Parziale A, Senatore R. Some observations on handwriting from a motor learning perspective. CEUR Workshop Proc. 2013;1022:6–10.
35. Deng L, Acero A, Bazzi I. Tracking vocal tract resonances using a quantized nonlinear function embedded in a temporal constraint. IEEE Trans Audio, Speech Lang Process. 2006;14(2):425–34.
36. Rabiner LR. Digital Processing of Speech Signal. Prentice - Hall; 1978.
37. Schroeder MR. Determination of the Geometry of the Human Vocal Tract by Acoustic Measurements. J Acoust Soc Am [Internet]. 1967;41(5):1283–94. Available from: https://doi.org/10.1121/1.1910429.
38. Atal BS, Chang JJ, Mathews M V., Tukey JW. Inversion of articulatory- to- acoustic transformation in the vocal tract by a computer-sorting technique. J Acoust Soc Am [Internet]. 1978;63(5):1535–55. Available from: https://doi.org/10.1121/1.381848.
39. Gómez-Vilda P, Gómez-Rodellar A, Vicente JMF, Mekyska J, Palacios-Alonso D, Rodellar-Biarge V, et al. Neuromechanical modelling of articulatory movements from surface electromyography and speech formants. Int J Neural Syst. 2019;29(02):1850039.
40. Gómez-Vilda P, Ferrández-Vicente JM, Rodellar-Biarge V. Simulating the phonological auditory cortex from vowel representation spaces to categories. Neurocomputing. 2013;114:63–75.
41. Gómez-Vilda P, Ferrández-Vicente JM, Rodellar-Biarge V, Álvarez-Marquina A, Mazaira-Fernández LM, Martínez Olalla R, et al. Neuromorphic detection of speech dynamics. Neurocomputing. 2011;74(8):1191–202.
42. Gómez-Vilda P, Ferrández-Vicente JM, Rodellar-Biarge V, Fernández-Baíllo R. Time-frequency representations in speech perception. Neurocomputing. 2009;72(4–6):820–30.
43. Boersma, Paul & Weenink D. Praat: doing phonetics by computer [Internet]. 2019. Available from: http://www.praat.org/.
44. Dromey C, Jang GO, Hollis K. Assessing correlations between lingual movements and formants. Speech Commun [Internet]. 2013;55(2):315–28. Available from: http://dx.doi.org/10.1016/j.specom.2012.09.001.
45. Gómez P, Mekyska J, Gómez A, Palacios D, Rodellar V, Álvarez A. Characterization of Parkinson's disease dysarthria in terms of speech articulation kinematics. Biomed Signal Process Control [Internet]. 2019;52:312–20. Available from: https://doi.org/10.1016/j.bspc.2019.04.029.
46. Gómez-Vilda P, Londral ARM, Rodellar-Biarge V, Ferrández-Vicente JM, de Carvalho M. Monitoring amyotrophic lateral sclerosis by biomechanical modeling of speech production. Neurocomputing [Internet]. 2015;151(P1):130–8. Available from: https://doi.org/10.1016/j.neucom.2014.07.074.
47. Hillenbrand J, Getty LA, Clark MJ, Wheeler K. Acoustic characteristics of American English vowels. J Acoust Soc Am [Internet].

1995;97(5):3099–111. Available from: http://asa.scitation.org/doi/10.1121/1.411872.
48. Pätzold M, Simpson AP. Acoustic analysis of German vowels in the Kiel Corpus of Read Speech. Arbeitsberichte des Instituts für Phonetik und Digit Sprachverarbeitung Univ Kiel [Internet]. 1997;32(1978):215–47. Available from: http://www.ipds.uni-kiel.de/kjk/pub_exx/aipuk32/mpas.pdf.
49. Whitfield J, Dromey C, Palmer P. Examining acoustic and kinematic measures of articulatory working space: effects of speech intensity. J Speech, Lang Hear Res. 2018;61(May):1–14.
50. Kuberski SR, Gafos AI. The speed-curvature power law in tongue movements of repetitive speech. PLoS ONE. 2019;14(3):1–25.
51. Li Deng, Xiaodong Cui, Pruvenok R, Huang J, Momen S, Yanyi Chen et al. A database of vocal tract resonance trajectories for research in speech processing. 2006;I-369-I–372.
52. Garofolo JS, Lamel LF, Fisher WM, Fiscus JG, Pallett DS, Dahlgren NL. TIMIT acoustic-phonetic continuous speech corpus LDC93S1. Philadelphia: Linguistic Data Consortium; 1993.
53. Barry WJ, Putzer M. Saarbruecken Voice Database [Internet]. Institute of Phonetics, Univ. of Saarland; Available from: http://www.stimmdatenbank.coli.uni-saarland.de/.
54. Godino-Llorente JI, Gomez-Vilda P, Blanco-Velasco M. Dimensionality reduction of a pathological voice quality assessment system based on Gaussian mixture models and short-term cepstral parameters. IEEE Trans Biomed Eng. 2006;53(10):1943–53.
55. Willems L. Robust formant analysis. IPO Rep. 1986;529:1–25.
56. Childers DG. Modern spectrum analysis. IEEE Press; 1978. p. 252–255.
57. Press WH, Teukolsky SA, Vetterling WT, Flannery BP. Numerical recipes in C: The art of scientific computing. 2nd ed. Cambridge University Press 1992.
58. Hogg RV, Ledolter J. Engineering Statistics. New York: MacMillan; 1987.
59. Bhattacharya U, Plamondon R, Dutta Chowdhury S, Goyal P, Parui SK. A sigma-lognormal model-based approach to generating large synthetic online handwriting sample databases. Int J Doc Anal Recognit. 2017;20(3):155–71.